\documentclass[pra,superscriptaddress,longbibliography,twocolumn]{revtex4-2}
\usepackage[utf8]{inputenc}
\usepackage[english]{babel}
\usepackage{amsmath}
\usepackage{amsfonts}
\usepackage{amssymb}
\usepackage{graphicx}
\usepackage{amsthm}
\usepackage{color}
\usepackage{hyperref}
\usepackage[caption=false]{subfig}
\usepackage{graphicx,float}
\usepackage{dcolumn}
\usepackage{bm}
\usepackage{mathrsfs}
\usepackage{txfonts}
\usepackage{CJK}
\usepackage[amssymb]{SIunits}
\usepackage{epsfig}
\usepackage{epstopdf}
\usepackage{lipsum}
\usepackage{placeins}
\usepackage{physics}

\newenvironment{SChinese}{%
\CJKfamily{gbsn}%
\CJKtilde
\CJKnospace}{}
\allowdisplaybreaks[2]

\begin{document}

\begin{CJK}{UTF8}{}
\begin{SChinese}

\title{Single-photon transport in a whispering-gallery mode microresonator directionally coupled with a two-level quantum emitter}

\author{Jiangshan Tang (唐江山)}  %
 \thanks{These authors contributed equally to this work.}
 \affiliation{College of Engineering and Applied Sciences, National Laboratory of Solid State Microstructures, and Collaborative Innovation Center of Advanced Microstructures, Nanjing University, Nanjing 210023, China}
  \affiliation{School of Physics, Nanjing University, Nanjing 210023, China}

\author{Lei Tang (唐磊)}  %
 \thanks{These authors contributed equally to this work.}
 \affiliation{College of Engineering and Applied Sciences, National Laboratory of Solid State Microstructures, and Collaborative Innovation Center of Advanced Microstructures, Nanjing University, Nanjing 210023, China}

\author{Keyu Xia (夏可宇)}  %
 \email{keyu.xia@nju.edu.cn}
    \affiliation{College of Engineering and Applied Sciences, National Laboratory of Solid State Microstructures, and Collaborative Innovation Center of Advanced Microstructures, Nanjing University, Nanjing 210023, China}
  \affiliation{School of Physics, Nanjing University, Nanjing 210023, China}


\begin{abstract}
We investigate the single-photon transport problem in the system of a Whispering-Gallery mode microresonator directionally coupled with a two-level quantum emitter (QE). This QE-microresonator coupling system can usually be studied by cavity quantum electrodynamics and the single-photon transport methods. However, we find that if we treat a two-level QE as a single-photon phase-amplitude modulator, we can also deal with such systems using the transfer matrix method. Further, in theory, we prove that these three methods are equivalent. The corresponding relations of respective parameters among these approaches are precisely deduced. Our work can be extended to a multiple-resonator system interacting with two-level QEs in a chiral way. Therefore, the transfer matrix method may provide a convenient and intuitive form for exploring more complex chiral QE-resonator interaction systems.
\end{abstract}

\maketitle

\end{SChinese}
\end{CJK}

\section{Introduction}\label{sec:intro}
The interaction of light and matter at the single-quantum level is the basis of essential physics of many phenomena and applications \cite{haroche2006exploring}, which has been extensively explored in various quantum systems, such as the single-mode waveguide coupling to quantum emitters (QEs) ~\cite{PhysRevA.76.062709,PhysRevA.78.063832,PhysRevA.79.023837,nature.566.7744,PhysRevA.101.053802,pucher2021atomic}, the Fabry-Perot cavity~\cite{PhysRevLett.101.100501,PhysRevLett.123.233604,ncomun.10.1038.2389,PhysRevApplied.15.064020,PRJ.413286} and the Whispering-Gallery mode (WGM) microresonator~\cite{science.319.5866,PhysRevLett.102.083601,PhysRevA.79.023838,PhysRevLett.110.213604,PhysRevLett.126.233602}.
In recent years, an emerging field of research called ``chiral quantum optics" \cite{nature21037}, in which the light-matter coupling is direction-dependent, has received extensive attention in the field of quantum nonreciprocity \cite{science.aaj2118} and exhibited chiral interactions of light and QEs \cite{PhysRevA.90.043802,PhysRevA.96.053804,science.aaq0327,PhysRevA.99.043833,OPTICA.393035,PRJ.405246,mehrabad2021chiral}.

To realize the chiral light-matter interaction, an external magnetic field is usually required to induce the magneto-optical effect \cite{APL.5.0057558} or make the energy of the QE undergo a Zeeman splitting \cite{nnano.2015.159}. It greatly limits the miniaturization and integration of single-photon devices. Recently, an all-optical approach, based on valley-selective response in transition metal dichalcogenides, has been described to overcome the limitations of magnetic materials \cite{ncommun.12.1.3746}. Towards on-chip chiral single-photon interfaces, non-magnetic schemes have been proposed based on a WGM microresonator chirally coupled with a two-level QE \cite{PhysRevA.90.043802,PhysRevX.5.041036,science.aaj2118,nature21037,PhysRevA.99.043833}.

Theoretically, the single-photon transport problem in the system of a WGM microresonator coupled to a waveguide can be solved by methods such as the cavity quantum electrodynamics (CQED) \cite{PhysRevA.57.R2293,JPBA.38.9.2005,nature05147,PhysRevA.75.023814,nature06274,PhysRevLett.99.173603}, and the single-photon transport (SPT) \cite{OL.30.002001,PhysRevLett.95.213001,PhysRevA.79.023837,PhysRevA.79.023838,Zheng2013}, and the transport matrix (TM) \cite{book.3540687866}.
Further, the WGM microresonator system containing a two-level QE has also been discussed under the framework of CQED and SPT theory \cite{PhysRevA.70.053808,science.319.5866,PhysRevA.79.023837,PhysRevA.79.023838}, even extending to the chiral interactions
\cite{PhysRevA.90.043802,science.aaj2118,nature21037,PhysRevA.99.043833}. However, how to deal with the chiral interaction of a WGM microresonator with a two-level QE using the TM method is still not available. And the inner link among these three methods also remain to be revealed.

In this paper, we demonstrate that the CQED, the SPT, and the TM methods are equivalent in dealing with the single-photon transport problem in a chiral QE-microresonator system. In the TM method, the effect of a two-level QE can be regarded as a single-photon phase-amplitude modulator. By introducing a nonlinear coefficient into the transfer relation, we can use the TM method to solve the single-photon transmission. Furthermore, we present the corresponding relations for the parameters among the three methods.
\section{System and model}\label{sec:model}
\begin{figure}
  \centering
  \includegraphics[width=0.8\linewidth]{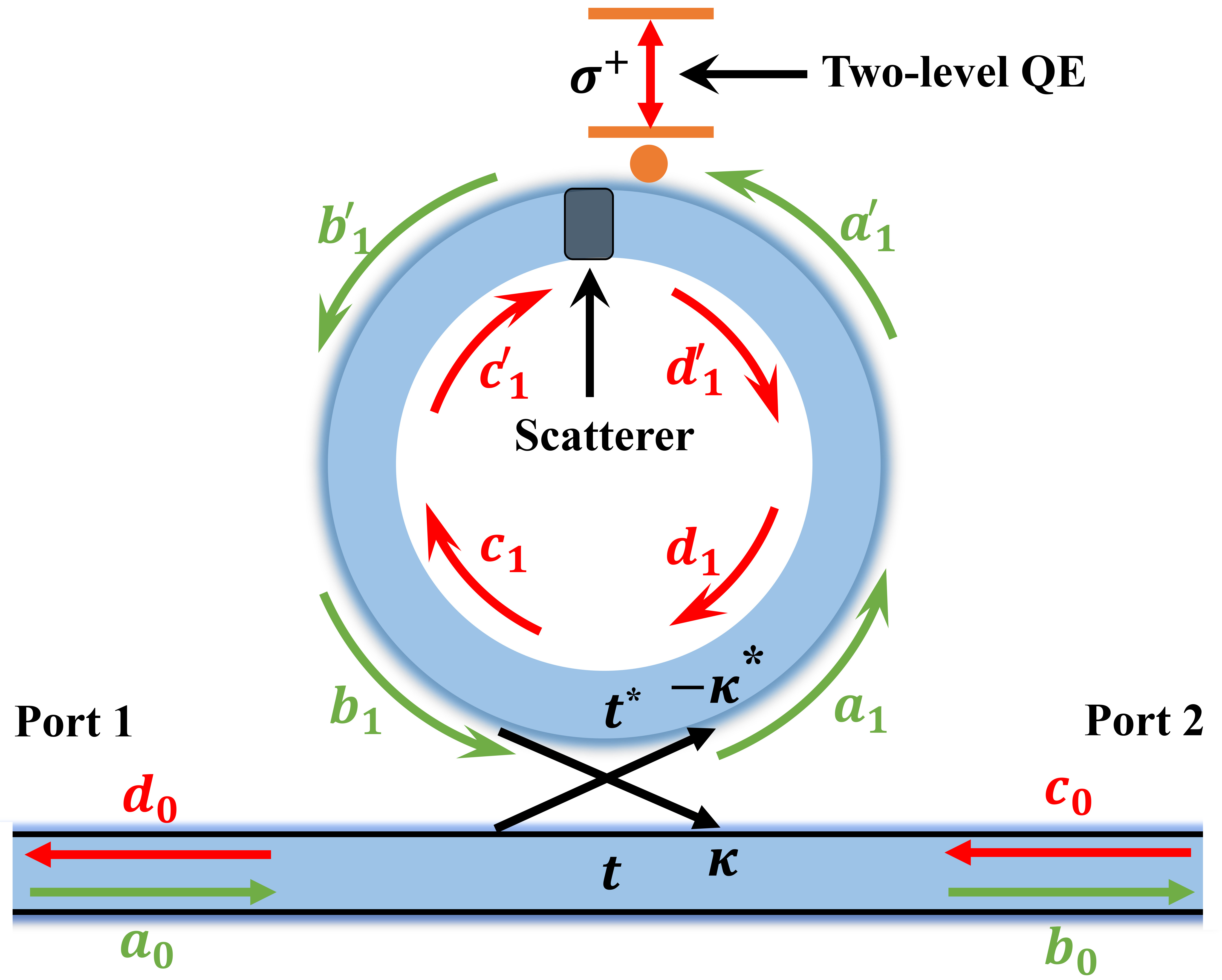} \\
\caption{Schematic of a chiral QE-microresonator system. A two-level QE is coupled to a WGM microresonator in a chiral way to form the QE-microresonator system. A waveguide is side coupled to the microresonator as input and output ports. A scatterer on the microresonator is considered to introduce backscattering. The arrows represent the propagating direction of a single photon for an input to the port 1 (green) or port 2 (red).}
\label{fig:FIG1}
\end{figure}

Schematic diagram of the system studied in this paper is shown in Fig.~\ref{fig:FIG1}. A WGM microresonator, which can be made with various material platforms, such as silicon oxynitride \cite{OE.15.017273,OL.33.002389}, ploymers \cite{OL.31.000456}, or silicon on insulator \cite{nphoton.2006.42,nphoton.2013.274,fmats.2015.00034}, simultaneously couples to a two-level QE and a waveguide to form a QE-microresonator system. A microresonator, with a radius $R$, supports two optical WGMs propagating in either clockwise (CW) or counterclockwise (CCW) direction. The evanescent fields of the WGMs are almost perfectly circular polarized with its polarization locked to the propagation direction and thus possesse optical chirality.
In practice, backscattering usually needs to be considered due to the surface roughness of microresonator. In this paper, we treat the backscattering as a scatterer
~\cite{nphys2063}, as shown in the black mark in Fig.~\ref{fig:FIG1}.

In our system, a two-level QE is chirally coupled to the microresonator. The chiral interaction between the QE and the evanescent field of WGM can be achieved by initializing QE in a specific spin ground state or shifting the transition energy with a polarization-selective optical Stark effect \cite{PhysRevA.90.043802,PhysRevX.5.041036,science.aaj2118,nature21037,ncomun.10.1038.2389,PhysRevA.99.043833,pucher2021atomic}. Here, we assume that the QE only interacts with the CCW WGM, because the evanescent field of the CCW mode is $\sigma^{+}$ polarized and only the $\sigma^{+}$-polarized transition of the prepared QE is allowed. Note that the QE-microresonator coupling strength is dependent on the propagating direction of light in our chiral system. In the forward case, the incident light from port $1$ excites the CCW mode in the microresonator, and it strongly couples with the QE. In the backward (port-$2$ incident) case, the CW mode is decoupled from the QE and thus the coupling strength $g$ is negligible (i.e., $g \approx 0$).
One can use a precisely positioned atom \cite{science.aaj2118,science.1237125}, a quantum dot (QD) \cite{RevModPhys.90.031002,science.aaq0327,nnano.2015.159,s41377-020-0244-9}, or a nanopillar covered by monolayers \cite{nnano.2015.75,ncomms15053,ncomms15093} to construct the two-level QE.

Below, we first provide the CQED, the SPT, and the TM methods to solve the response of the system. Then we show these three methods are equivalent if we treat the two-level QE as a single-photon phase-amplitude modulator. We only discuss the forward case ($g \ne 0$) in detail, and the backward case corresponds to the system without a QE ($g = 0$).

\subsection{Cavity Quantum Electrodynamics Method}\label{subsec:CQED}
In this section, we discuss the CQED method to solve our model. For a coupled atom-microresonator system, it has been analyzed in Ref.~\cite{science.319.5866}. Here, we discuss the chiral coupling using the same approach. We consider a two-level QE with transition frequency $\omega_{\text{qe}}$ is coupled to the CCW mode and decoupled to the opposite mode. The two degenerate WGMs, with same resonant frequency $\Omega$ and dissipation $\kappa_{\text{tol}}$, are assumed to be coupled with each other in a strength $h$ due to the scatterer. Here, we divide the dissipation $\kappa_{\text{tol}}$ into two parts, the intrinsic decay rate of $\kappa_{\text{in}}$ and the external loss of $\kappa_{\text{ex}}$, satisfying $\kappa_{\text{tol}}=\kappa_{\text{in}}+\kappa_{\text{ex}}$.  A weak coherent field of frequency $\omega$ with a amplitude $\alpha_{\text{in}}$ drives the CCW mode $a$. To a good single-photon approximation, $\alpha_{\text{in}}\ll1$. In a frame rotating at the frequency $\omega$, the Hamiltonian of our system can be obtained \cite{scully1999quantum}
\begin{equation}\label{eq:Hamiltonian1}
  \begin{aligned}
H=&-\Delta_{1} a^{\dagger} a-\Delta_{2} \sigma^{+} \sigma^{-}-\Delta_{1} b^{\dagger} b+i \sqrt{2 \kappa_{\text{ex}}}\alpha_{\text{in}}\left( a^{\dagger}- a\right) \\
&+g\left(a^{\dagger} \sigma^{-}+\sigma^{+} a\right)+h\left(a^{\dagger} b+b^{\dagger} a\right)\;,
\end{aligned}
\end{equation}
where $g$ represents the coupling strength between the CCW mode and the QE. $\Delta_1=\omega-\Omega$ and $\Delta_2=\omega-\omega_{\text{qe}}$ are the detunings.  $b$ is the annihilation operator of the CW mode. $\sigma^{\pm}$ are the raising and lowering operators describing the two-level QE. It is worth noting that if we consider the coupling of two microresonators instead of the scatterer, the Hamiltonian has the same form as Eq.~(\ref{eq:Hamiltonian1}). In this case, $h$ describes the coupling strength between the two microresonators. Introducing the dissipation of the QE, $\gamma$, the evolution of the system can be found by solving the master equation,
\begin{equation}\label{eq:mastereq}
  \begin{aligned}
\dot{\rho}=&-\mathrm{i}[H, \rho]+\kappa_{\text{tol}}\left(2 a \rho a^{\dagger}-a^{\dagger} a \rho-\rho a^{\dagger} a\right)\\
&+\kappa_{\text{tol}}\left(2 b \rho b^{\dagger}-b^{\dagger} b \rho-\rho b^{\dagger} b\right) \\
&+\gamma\left(2 \sigma^{-} \rho \sigma^{+}-\sigma^{+} \sigma^{-} \rho-\rho \sigma^{+} \sigma^{-}\right)\;,
\end{aligned}
\end{equation}
where $\rho$ is the density operator. From Eq.~(\ref{eq:mastereq}), we can derive the equations of motion,
\begin{subequations}\label{eq:motion}
  \begin{align}
    &\dot{a}=i\tilde{\Delta}_1a+\alpha_{\text{in}}\sqrt{2\kappa_{\text{ex}}}-ig\sigma^{-}-ihb \;, \\
   &\dot{\sigma}^{-}=i\tilde{\Delta}_2\sigma^{-}+ig\sigma_{z}a\;,\\
  &\dot{b}=i\tilde{\Delta}_1b-iha\;,
  \end{align}
  \end{subequations}
and have the steady-state solution
\begin{equation}\label{eq:steady}
  a=\frac{i \alpha_{\text{in}} \sqrt{2 \kappa_{\text{ex}}}\tilde{\Delta}_1\tilde{\Delta}_2}{\tilde{\Delta}_1\left(\tilde{\Delta}_1\tilde{\Delta}_2+\sigma_{z} g^{2}\right)-\tilde{\Delta}_2 h^{2}}\;,
\end{equation}
where, $\sigma_{z}=\sigma^{+}\sigma^{-}-\sigma^{-}\sigma^{+}$, $\tilde{\Delta}_1=\Delta_1+i\kappa_{\text{tol}}$ and $\tilde{\Delta}_2=\Delta_2+i\gamma$. For simplicity, we have omitted the notation $\langle\cdots\rangle$. According to the input-output relation, $a_{\text{out}}=\alpha_{in}-\sqrt{2\kappa_{\text{ex}}}a$, we can get the transmission amplitude in the port 2:
\begin{equation}\label{eq:transmission5}
  t_{\omega}=\frac{\tilde{\Delta}_1\left[\left(\Delta_{1}+i \kappa_{\text{in}}-i \kappa_{\text{ex}}\right)\tilde{\Delta}_2+\sigma_{z} g^{2}\right]-\tilde{\Delta}_2 h^{2}}{\tilde{\Delta}_1\left(\tilde{\Delta}_1\tilde{\Delta}_2+\sigma_{z} g^{2}\right)-\tilde{\Delta}_2 h^{2}}
\end{equation}
The transmission of port 2 can be obtained from $T=|t_{\omega}|^2$.

Moreover, the fully quantum dynamics of the system can be solved by a numerical solution to the master Eq.~(\ref{eq:mastereq}) using truncated space of photon number for the WGMs.

\subsection{Single-Photon Transport Method}\label{subsec:SPT}
Hereafter, we consider only a single photon in our system. Based on the single-photon transport theory \cite{OL.30.002001,PhysRevA.79.023837,PhysRevA.79.023838}, our previous work has given a transmission amplitude for the system interesting in this paper, see Eq.~(15) in Ref.~\cite{PhysRevA.99.043833}. Since the QE is decoupled to the CW  mode in our system, the form of the transmission amplitude reads:
\begin{equation}\label{eq:transmission6}
  t_{\omega}=\frac{\tilde{\Delta}_1\left[\left(\Delta_{1}+i \kappa_{\text{in}}-i \kappa_{\text{ex}}\right)\tilde{\Delta}_2- g^{2}\right]-\tilde{\Delta}_2 h^{2}}{\tilde{\Delta}_1\left(\tilde{\Delta}_1\tilde{\Delta}_2- g^{2}\right)-\tilde{\Delta}_2 h^{2}}\;.
\end{equation}

If we consider $\sigma_z=-1$ in the CQED method, that is, the weak probe field approximation \cite{scully1999quantum}, we can find that the Eq.~(\ref{eq:transmission5}) and the Eq.~(\ref{eq:transmission6}) are equivalent. In the Sec.~\ref{subsec:phaser}, we will verify
that the TM method is consistent with the SPT method.

\subsection{Transport Matrix Model}\label{subsec:TM}
Next, we study the chiral QE-microresonator system using the TM method. Under the notation in Fig.~\ref{fig:FIG1}, the coupling relation between the waveguide and the microresonator can be written as
\begin{equation}\label{eq:rela1}
  \left\{\begin{array} { l }
{ a _ { 1 } = t ^ { * } b _ { 1 } - \kappa ^ { * } a _ { 0 } } \\
{ b _ { 0 } = t a _ { 0 } + \kappa b _ { 1 } }
\end{array}\;, \quad \left\{\begin{array}{l}
c_{1}=t^{*} d_{1}-\kappa^{*} c_{0} \\
d_{0}=t c_{0}+\kappa d_{1}
  \end{array}\right.\right.\;,
\end{equation}
where $t$ and $\kappa$ are the transmission and the coupling coefficients, and $|t|^2+|\kappa|^2=1$ for lossless coupling. Written in a matrix form:
\begin{equation}\label{eq:tm1}
  \left(\begin{array}{l}
a_{0} \\
b_{0} \\
c_{0} \\
d_{0}
\end{array}\right)=\frac{1}{\kappa^{*}}\left(\begin{array}{cccc}
-1 & t^{*} & 0 & 0 \\
-t & 1 & 0 & 0 \\
0 & 0 & -1 & t^{*} \\
0 & 0 & -t & 1
\end{array}\right)\left(\begin{array}{l}
a_{1} \\
b_{1} \\
c_{1} \\
d_{1}
\end{array}\right)\equiv M_{\text{cpl}}\left(\begin{array}{l}
a_{1} \\
b_{1} \\
c_{1} \\
d_{1}
\end{array}\right).
\end{equation}

The size of the QE and the scatterer is much smaller than the structure of microresonator, so theoretically they can be treated as particles. Without loss of generality, we assume that the QE and the scatterer are in the same position, see Fig.~\ref{fig:FIG1}. Thus, the coupling points of the waveguide, the QE and the microresonator divide the microresonator into two parts with lengths $L_1$ and $L_2$, satisfying $L_1+L_2=2\pi R$. The field component notations are shown in Fig.~\ref{fig:FIG1}. When a single photon propagates around the microresonator, it will accumulate propagation phases $\theta_{j}=\beta L_j$, and may attenuate with loss $\alpha_{j}(L_j)$ ($j = 1,2$) \cite{book.3540687866}. We take $\theta=\theta_1+\theta_2$ and $\alpha=\alpha_1\alpha_2$.
The factor $\beta$ is the propagation constant in the microresonator as given by $\beta=n_{\text{eff}}\omega/c$, where $n_{\text{eff}}$ is the effective refractive index and $\omega$ is the frequency. Thus, we have the transfer relation
\begin{subequations}\label{eq:tm2}
  \begin{align}
    &~~~~~~~~~~~~~~~~~~~~~~~~~~~~\left(\begin{array}{l}
a_{1} \\
b_{1} \\
c_{1} \\
d_{1}
\end{array}\right)=M_{\text{pro}}\left(\begin{array}{l}
a_{1}^{\prime} \\
b_{1}^{\prime} \\
c_{1}^{\prime} \\
d_{1}^{\prime}
\end{array}\right) \;, \\
   &M_{\text{pro}}=\left(\begin{array}{cccc}
\alpha_{1}^{-1} e^{-i \theta_{1}} & 0 & 0 & 0 \\
0 & \alpha_{2} e^{i \theta_{2}} & 0 & 0 \\
0 & 0 & \alpha_{2}^{-1} e^{-i \theta_{2}} & 0 \\
0 & 0 & 0 & \alpha_{1} e^{i \theta_{1}}
\end{array}\right)\left(\begin{array}{c}
a_{1}^{\prime} \\
b_{1}^{\prime} \\
c_{1}^{\prime} \\
d_{1}^{\prime}
\end{array}\right) \;.
  \end{align}
\end{subequations}
We refer to $M_{\text{cpl}}$ and $M_{\text{pro}}$ as coupling and propagation matrices. Combining Eqs.~(\ref{eq:tm1}) and (\ref{eq:tm2}), we obtain the transfer matrix as
\begin{equation}\label{eq:tm3}
  \left(\begin{array}{l}
a_{0} \\
b_{0} \\
c_{0} \\
d_{0}
\end{array}\right)=M_{\text{cpl}}M_{\text{pro}}\left(\begin{array}{l}
a_{1}^{\prime} \\
b_{1}^{\prime} \\
c_{1}^{\prime} \\
d_{1}^{\prime}
\end{array}\right) \;.
\end{equation}

We consider a single input of port 1 ($c_0=0$). It excites the CCW-direction WGM. In the following, we will discuss four cases:
\\
1. No two-level QE and No scatterer

We first consider the case without two-level QEs and scatterers, the transfer relation of the field amplitudes in the microresonator can be directly obtained,  $a_1^{'}$=$b_1^{'}$ and $c_1^{'}$=$d_1^{'}$. In the absence of scatterers, the CCW and CW modes are decoupled. We can get the transmission amplitude in the port 2 \cite{book.3540687866},
\begin{equation}\label{eq:transmission1}
t_{\omega}=\frac{b_{0}}{a_{0}}=\frac{-t+\alpha e^{i \theta}}{-1+\alpha t^{*} e^{i \theta}}\;.
\end{equation}
\\
2. No two-level QE and Consider scatterer

In this case, we consider the effect of the scatterer in the microresonator. The relation between the amplitudes can be written as $b_1^{'}=t_sa_1^{'}+r_sc_{1}^{'}$ and $d_1^{'}=t_sc_1^{'}+r_sa_{1}^{'}$. $t_s$/$r_s$ are the transmission/reflection coefficients introduced by the scatterer, and they satisfy $|t_s|^2+|r_s|^2=1$ when the dissipation of the scatterer is neglected. The two WGMs are coupled to each other in this case. We assume the scatterer is weak, thus we can write $t_s/r_s$ in the following form \cite{nphys2063}:
\begin{equation}\label{eq:tm3}
t_s=\text{cos}\epsilon\approx1-\frac{\epsilon^2}{2}, \quad
r_s=i\text{sin}\epsilon\approx i\epsilon\;.
\end{equation}
Then we have the transmission amplitude in the port 2,
\begin{equation}\label{eq:transmission2}
 t_{\omega}=\frac{b_{0}}{a_{0}}=\frac{-t+\alpha e^{i \theta} \frac{t_{s}-t^{*} \alpha e^{i \theta}}{1-t_{s} t^{*} \alpha e^{i \theta}}}{-1+\alpha t^{*} e^{i \theta} \frac{t_{s}-t^{*} \alpha e^{i \theta}}{1-t_{s} t^{*} \alpha e^{i \theta}}}\;.
\end{equation}
\\
3. Single two-level QE and No scatterer

Here, we study the effect of a two-level QE directionally coupled to a microresonator. Because the QE is in a specific spin ground state or the polarization-selective energy level transition, the coupling of the QE and the evanescent field on the microresonator is direction-dependent. The reflection of single-photon propagation will vanish due to such chiral QE-light interaction \cite{PhysRevA.90.043802,PhysRevA.99.043833}. In this case, the single photon will not excite the CW mode, leading to the decoupling between the CCW and CW modes. We assume the single photon through the two-level QE with a transmission coefficient $t_{\text{qe}}$, i.e., $b_1^{'}=t_{\text{qe}}a_1^{'}$ and $d_1^{'}=c_1^{'}$, such that
\begin{equation}\label{eq:transmission3}
  t_{\omega}=\frac{b_{0}}{a_{0}}=\frac{-t+\alpha e^{i \theta}t_{\text{qe}}}{-1+\alpha t^{*} e^{i \theta}t_{\text{qe}}}\;.
\end{equation}
The specific form of $t_{\text{qe}}$ will be discussed below.\\

4. Single two-level QE and Consider the scatterer

Combining with the above discussions, we can get the transfer relation of the field amplitudes, with considering both a two-level QE directionally coupled to the microresonator and a scatterer, $b_1^{'}=t_{\text{qe}}\left(t_sa_1^{'}+r_sc_1^{'}\right)$ and $d_1^{'}=t_sc_1^{'}+r_sa_1^{'}$.  The transmission amplitude can be calculated as
\begin{equation}\label{eq:transmission4}
   t_{\omega}=\frac{b_{0}}{a_{0}}=\frac{-t+\alpha e^{i \theta}t_{\text{qe}} \frac{t_{s}-t^{*} \alpha e^{i \theta}}{1-t_{s} t^{*} \alpha e^{i \theta}}}{-1+\alpha t^{*} e^{i \theta}t_{\text{qe}} \frac{t_{s}-t^{*} \alpha e^{i \theta}}{1-t_{s} t^{*} \alpha e^{i \theta}}}\;.
\end{equation}

\subsection{Single-Photon Phase-amplitude Modulator}\label{subsec:phaser}
We define the round-trip time of microresonator, $\tau_{\text{rt}}=2\pi Rn_{\text{eff}}/c$, that a photon needs to make a round trip in the microresonator of length $2\pi R$. It is the inverse of the free spectral range $\mathcal{F}$, i.e., $\tau_{\text{rt}}=1/\mathcal{F}$ \cite{wolfgramm2012}. Since $\mathcal{F}\gg1$ for a microresonator, $\tau_{\text{rt}}$ is a small amount. We have $\text{exp}(i\theta)=\text{exp}\left[i\left(\omega-\Omega\right)\tau_{\text{rt}}\right]\approx 1+i\Delta_1\tau_{\text{rt}}$. For a single photon having travelled a round trip in the microresonator, we can get $a_1(\tau_{\text{rt}})=\alpha t^{*}a_1(0)$ from the transfer relation. The circulating power meets $|a_1(\tau_{\text{rt}})|^2=\alpha^2 t^{2}|a_1(0)|^2$. On the other hand, we have $|a_1(\tau_{\text{rt}})|^2=\text{exp}(-2\kappa_{\text{tol}}\tau_{\text{rt}})|a_1(0)|^2$, from the dissipation of the microresonator. Hence, we obtain
\begin{subequations}\label{eq:para1}
  \begin{align}
    &\alpha=e^{-\kappa_{\text{in}}\tau_{\text{rt}}}\approx 1-\kappa_{\text{in}}\tau_{\text{rt}} \;, \\
   &t=e^{-\kappa_{\text{ex}}\tau_{\text{rt}}}\approx 1-\kappa_{\text{ex}}\tau_{\text{rt}} \;.
  \end{align}
  \end{subequations}

Because the size of the two-level QE is much smaller than the bend structure of the microresonator, the interaction between the evanescent field and the QE can be equivalent to that of a waveguide directionally coupled with a two-level QE \cite{PhysRevA.90.043802}, with a transmission coefficient
\begin{equation}\label{eq:tqe}
  t_{\text{qe}}=\frac{\omega-\omega_{\text{qe}}+i\left(\gamma-\Gamma\right)}{\omega-\omega_{\text{qe}}+i\left(\gamma+\Gamma\right)}.
\end{equation}
$\Gamma$ is the decay rate from the QE into the microresonator. Therefore, substituting the Eqs.~(\ref{eq:tm3}), (\ref{eq:para1}) and (\ref{eq:tqe}) into Eq.~(\ref{eq:transmission4}), and ignoring the second-order small quantity, we have
\begin{equation}\label{eq:transmission7}
  \begin{aligned}
&t_{\omega}=\frac{-t+\alpha e^{i \theta} t_{\text{qe}} \frac{t_{s}-t^{*} \alpha e^{i \theta}}{1-t_{s} t^{*} \alpha e^{i \theta}}}{-1+\alpha t^{*} e^{i \theta} t_{\text{qe}} \frac{t_{s}-t^{*} \alpha e^{i \theta}}{1-t_{s} t^{*} \alpha e^{i \theta}}}\\
\approx &
\frac{\kappa_{\text{ex}} \tau_{\text{rt}}-1+\left(1-\kappa_{\text{in}} \tau_{\text{rt}}+i \Delta_{1} \tau_{\text{rt}}\right)\left[\left(1-\frac{2 i \Gamma}{\tilde{\Delta}_{2}+i \Gamma}\right)\left(1+\frac{\varepsilon^{2}}{i \tilde{\Delta}_{1} \tau_{\text{rt}}+\varepsilon^{2} / 2}\right)\right]}{-1+\left(1+i \tilde{\Delta}_{1} \tau_{\text{rt}}\right)\left[\left(1-\frac{2 i \Gamma}{\tilde{\Delta}_{2}+i \Gamma}\right)\left(1+\frac{\varepsilon^{2}}{i \tilde{\Delta}_{1} \tau_{\text{rt}}+\varepsilon^{2} / 2}\right)\right]}\\
\approx & \frac{\tilde{\Delta}_{1}\left[\left(\Delta_{1}+i\kappa_{\text{in}}-i\kappa_{\text{ex}}\right)\tilde{\Delta}_{2}-\Gamma\left(2 / \tau_{\text{rt}}-\kappa_{\text{tol}}\right)\right]-\tilde{\Delta}_{2} \frac{\varepsilon^{2}}{\tau_{\text{rt}}^{2}}}{\tilde{\Delta}_{1}\left[\tilde{\Delta}_{1}\tilde{\Delta}_{2}-\Gamma\left(2 / \tau_{\text{rt}}-\kappa_{\text{tol}}\right)\right]-\tilde{\Delta}_{2} \frac{\varepsilon^{2}}{\tau_{\text{rt}}^{2}}}\\
=&
\frac{\tilde{\Delta}_{1}\left[\left(\Delta_{1}+i\kappa_{\text{in}}-i\kappa_{\text{ex}}\right)\tilde{\Delta}_{2}-\Gamma\left(2 \mathcal{F}-\kappa_{\text{tol}}\right)\right]-\tilde{\Delta}_{2} \left(\varepsilon\times\mathcal{F}\right)^2}{\tilde{\Delta}_{1}\left[\tilde{\Delta}_{1}\tilde{\Delta}_{2}-\Gamma\left(2 \mathcal{F}-\kappa_{\text{tol}}\right)\right]-\tilde{\Delta}_{2} \left(\varepsilon\times\mathcal{F}\right)^2}.
\end{aligned}
\end{equation}
Comparing Eq.~(\ref{eq:transmission7}) with Eq.~(\ref{eq:transmission6}), we can find that if we take
\begin{equation}\label{eq:para2}
\Gamma\left(2\mathcal{F}-\kappa_{\text{tol}}\right)=g^2, \quad
\varepsilon \times\mathcal{F}=h \;,
\end{equation}
then the TM method and the SPT method are consistent. This also proves that the assumption of Eq.~(\ref{eq:tqe}) is valid.

Therefore,  the two-level QE can be treated as a single-photon phase-amplitude modulator, that causes a change in propagation phase and an amplitude modulation when the single photon passes through it. We divide Eq.~(\ref{eq:tqe}) into two parts: $t_{\text{qe}}=\exp\left(i\varphi_{\text{pha}}\right) \exp\left(-\varphi_{\text{dis}}\right)$, where $\exp\left(i\varphi_{\text{pha}}\right)=\text{arg}\left(t_{\text{qe}}\right)$ represents the change of the phase and $\exp\left(-\varphi_{\text{dis}}\right)=|t_{\text{qe}}|$ describes the attenuation of the amplitude.
It is worth noting that the additional propagation phase introduced by the two-level QE can be equivalent to a change of the effective resonance frequency of the microresonator. When a single photon travels around the microresonator, in the absence of the QE,
we have $\theta=2\pi R\beta=2\pi m \omega/\Omega$, where $m=\Omega n_\text{eff}R/c$ is the modal number. But if we consider the chiral QE-microresonator interaction, the additional propagation phase $\varphi_{\text{pha}}$ leads to $\theta+\varphi_{\text{pha}}=2\pi m \omega/\Omega_{\text{eff}}$. The effective resonance frequency of the microresonator is
\begin{equation}\label{eq:Omegaeff}
  \Omega_{\text{eff}}\approx \Omega \left(1- \frac{\varphi_{\text{pha}}\Omega}{2\pi m\omega} \right)\;,
\end{equation}
and $\Omega_{\text{eff}}\approx \Omega\left[1-\varphi_{\text{pha}}/2\pi m\right]$ for $\Omega/\omega\approx1$.

In general, by equating a two-level QE directionally coupled with a microresonator to a single-photon phase-amplitude modulator, we can use the TM method to solve the single-photon transport problem in such chiral QE-microresonator systems. This only needs to be multiplied by a transmission coefficient $t_{\text{qe}}$ in the transfer relation. Further, this approach can be extended to the system in which multiple QEs are coupled to microresonators in a chiral way.

\begin{figure}
  \centering
  \includegraphics[width=1.0\linewidth]{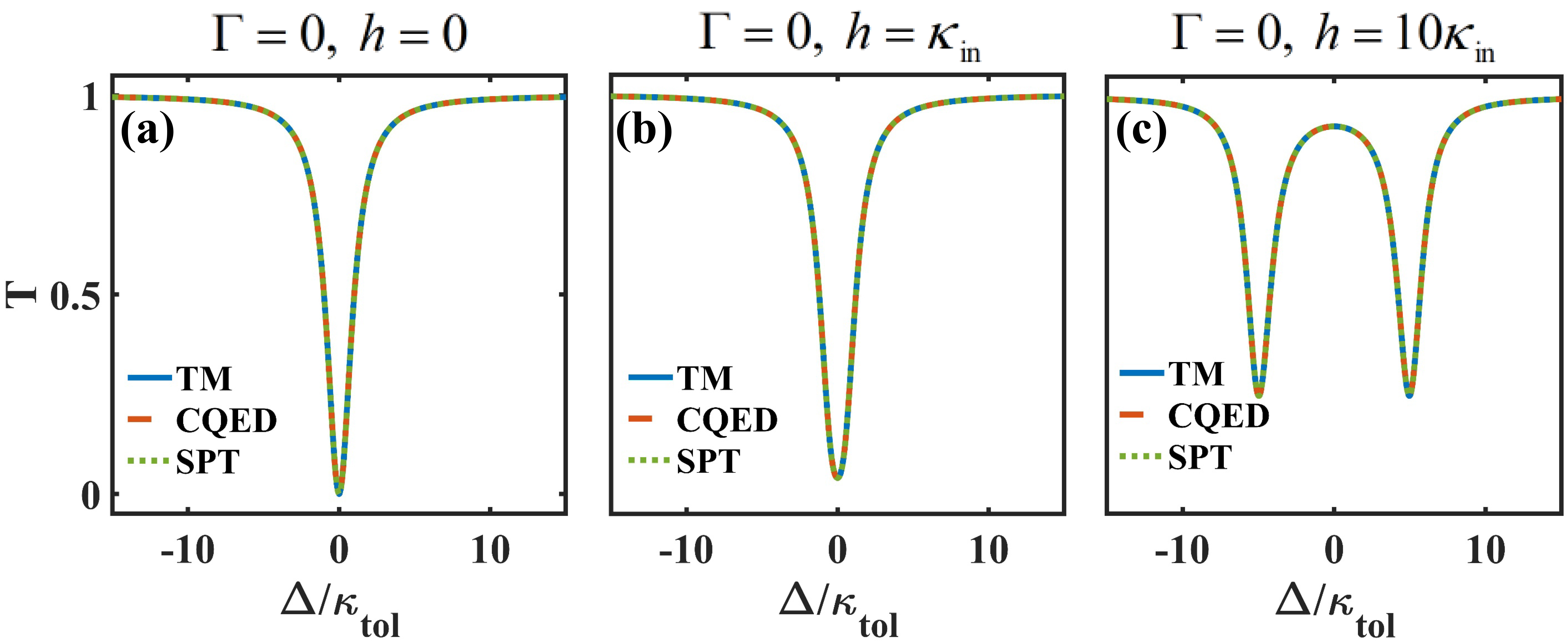} \\
\caption{The transmission spectra of a waveguide coupled with a microresonator. The blue solid, the red dashed, and the green dotted curves are calculated by the TM method, the CQED method and the SPT method, respectively. The setting in the following figures is the same. (a) In absence of backscattering. (b) and (c) In presence of the backscattering with strengths $h=\kappa_{\text{in}}$ and $h=10\kappa_{\text{in}}$, respectively. See the Sec.~\ref{sec:results} for other parameters.}
\label{fig:FIG2}
\end{figure}

\begin{figure}
  \centering
  \includegraphics[width=1.0\linewidth]{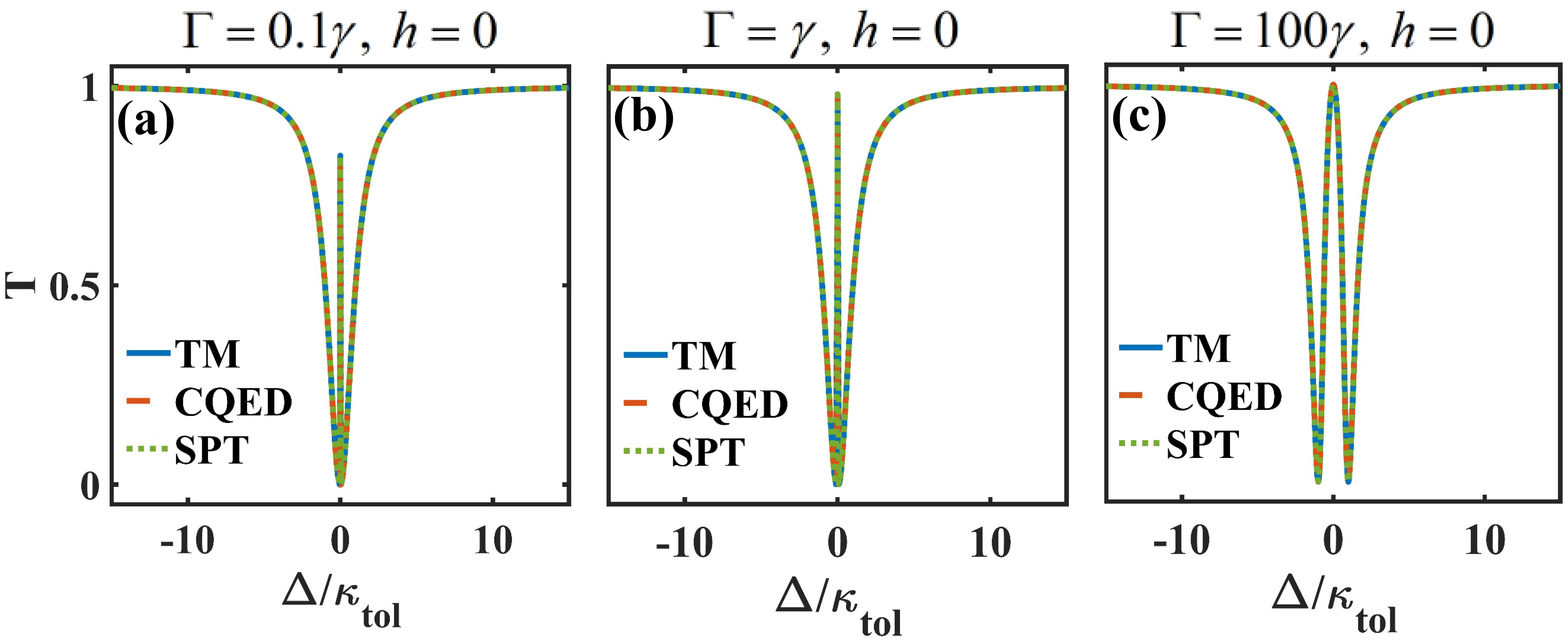} \\
\caption{The transmission spectra for a chiral QE-microresonator system without considering the backscattering. $\Gamma=0.1\gamma$, $\Gamma=\gamma$ and $\Gamma=100\gamma$ in (a-c), respectively.}
\label{fig:FIG3}
\end{figure}

\section{Results}\label{sec:results}
Below we numerically study our system to prove the consistency of the three methods. For the TM method and the SPT method, we solve the Eqs.~(\ref{eq:transmission4}) and (\ref{eq:transmission6}) directly, whereas for the CQED method, we perform a full quantum dynamics simulation using Eq.~(\ref{eq:mastereq}). We set a prepared QD as the two-level QE for coupling to a silicon-based microresonator in a chiral way. The experimentally available parameters are chosen as \cite{RevModPhys.90.031002,lpor.201100018,fmats.2015.00034}: $R = \text{10.5 }~\micro\meter$, $n_{\text{eff}}=1.5$, $\mathcal{F}/2\pi=3~\tera\hertz$, $\alpha_{\text{in}}=0.1$, and $\gamma/2\pi=6~\mega\hertz$. The conversion relationships between the parameters of the three methods are given by Eqs.~(\ref{eq:para1}) and (\ref{eq:para2}). We take $t=\alpha=0.99$, thus satisfying the critical coupling condition, $\kappa_{\text{ex}}/2\pi=\kappa_{\text{in}}/2\pi=30~\giga\hertz$. The frequency of the QD is resonant with the microresonator, i.e., $\omega_{\text{qe}}=\Omega$. We first consider the case without a two-level QD, corresponding to $\Gamma=0~(g=0)$, shown in Fig.~\ref{fig:FIG2}. When the strength $h=0$, the deep of transmission appears at the resonance point [see Fig.~\ref{fig:FIG2}(a)]. As the strength $h$ increases, the transmission spectrum gradually splits [see Fig.~\ref{fig:FIG2}(b-c)]. The calculation results of the three methods are exactly the same.

\begin{figure}
  \centering
  \includegraphics[width=1.0\linewidth]{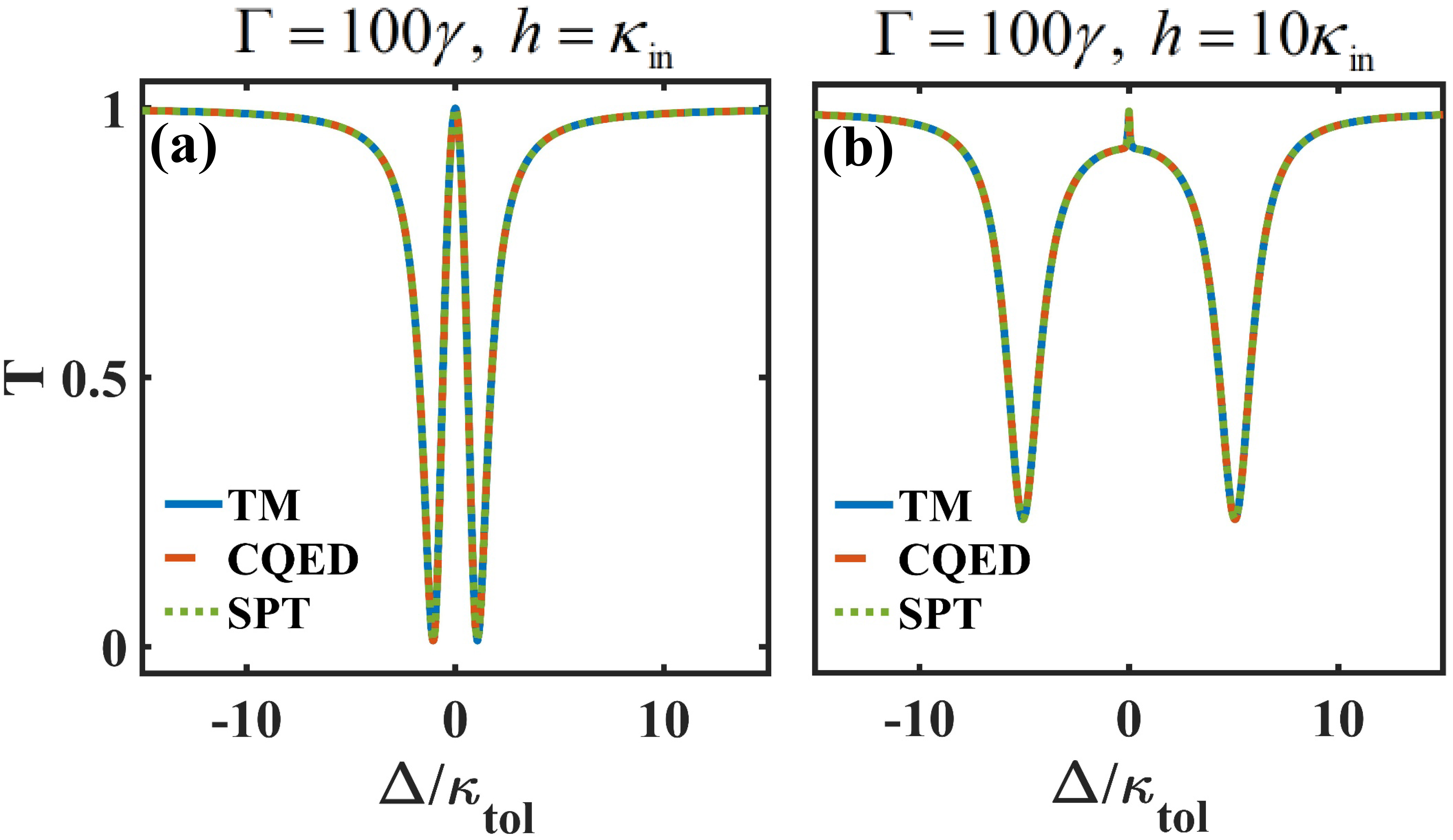} \\
\caption{The transmission spectra for a chiral QE-microresonator system. $\Gamma=100\gamma$, $h=\kappa_{\text{in}}$ in (a) and $\Gamma=100\gamma$, $h=10\kappa_{\text{in}}$ in (b).}
\label{fig:FIG4}
\end{figure}

Then we consider the chiral coupling of a two-level QD. By modeling the two-level QD chirally coupled to the microresonator as a single-photon phase-amplitude modulator, we can use the TM method to solve such problems. Figure~\ref{fig:FIG3} shows the transmission spectra without scatterers. The presence of the two-level QD causes the transmission spectrum to split \cite{PhysRevA.99.043833}. We can find that the transmission spectra calculated by the three methods are consistent regardless of whether it is under weak coupling, $\Gamma=0.1\gamma$ and $\Gamma=\gamma$ ($g/\kappa_{\text{tol}}=0.03$ and $g/\kappa_{\text{tol}}=0.1$ ), or strong coupling, $\Gamma=100\gamma$ ($g/\kappa_{\text{tol}}=1$). The results taking into account the effect of backscattering are shown in Fig.~\ref{fig:FIG4}. We consider the case of strong coupling, $\Gamma=100\gamma$. Whether it is in the case of weak backscatter [see Fig.~\ref{fig:FIG4}(a)] or strong backscatter [see Fig.~\ref{fig:FIG4}(b)], the calculation results are consistent. Therefore, our numerical results further confirm the above theoretical analyses and prove the correctness of the parameter relationships of these three methods.

\section{Conclusion}\label{sec:conc}
We demonstrate that a two-level QE can be treated as a single-photon phase-amplitude modulator in a chiral QE-microresonator system. Based on this, we can solve the single-photon transport problem by the method of TM. Theoretical analyses and numerical results confirm that the TM method is consistent with CQED and SPT methods. The conversion for the parameters of these three methods is explicitly derived. Without loss of generality, the TM method can be extended to solve the single-photon transport problem of any number of two-level QEs chirally coupled to multiple microresonators.

\section*{Acknowledgements}
This work was supported by the National Key R\&D Program of China (Grants No. 2019YFA0308700, No. 2017YFA0303703, No. 2017YFA0303701), the National Natural Science Foundation of China (Grant Nos. 11874212, 11890704, 61671279,11574145,11690031), the Fundamental Research Funds for the Central Universities (021314380095) and Program for Innovative Talents and Entrepreneurs in Jiangsu (Grant No. JSSCTD202138).

%

\providecommand{\noopsort}[1]{}\providecommand{\singleletter}[1]{#1}%

\end{document}